# Probabilistic Re-aggregation Algorithm

Alistair Reid, Xinyue Wang, Simon O'Callaghan, Daniel Steinberg, Lachlan McCalman

Spatial data about individuals or businesses is often aggregated over polygonal regions to preserve privacy, provide useful insight and support decision making. Given a particular aggregation of data (say into local government areas), the *re-aggregation* problem is to estimate how that same data would aggregate over a different set of polygonal regions (say electorates) without having access to the original unit records. Data61 is developing new re-aggregation algorithms that both estimate confidence intervals of their predictions and utilize additional related datasets when available to improve accuracy. The algorithms are an improvement over the current re-aggregation procedure in use by the ABS, which is manually applied by the data user, less accurate in validation experiments and provides a single best guess answer. The algorithms are deployed in an accessible web service that automatically learns a model and applies it to user-data. This report formulates the re-aggregation problem, describes Data61's new algorithms, and presents preliminary validation experiments.

## 1 Introduction

Unit-record datasets about individuals or businesses are often aggregated over polygonal regions to preserve privacy, provide insight and support decision making. To standardise aggregations, the Australian Bureau of Statistics (ABS) maintains a specification of spatial polygonal structures called the Australian Statistical Geography Standard (ASGS[3]). Many other structures typically used by government administration such as electoral divisions and local government areas are also supported. The structural *hierarchy* is shown in Figure 1, where a connection means that the right structure is *composed* of a union of the left class of structures.

Commonly, organisations need to convert existing data onto new geographic areas to understand the data for a specific region, or to allow for integration of their data with other data that was aggregated to a different level. While aggregations provide an interpretable summary of the data, their transformation from unit records is irreversible, because the spatial distribution of the original counts can not be recovered from the resulting tabular data.

We define *re-aggregation* as the process of predicting how the data would aggregate to an alternative spatial geometry, without assuming access to the original unit records. For example, if crime is reported by Local Government Area, a government agency may want statistics for particular State Electoral Divisions (SEDs). Equivalently, if crop yields are privately collected per farm, researchers may need to align the yield data with soil chemistry recorded on a square kilometre grid. Likewise, if the ABS reports counts of registered businesses by SA2, the treasury may ask for the data by Commonwealth Electoral Divisions (CED) to advise the House of Representatives. All of these scenarios are examples of re-aggregation.

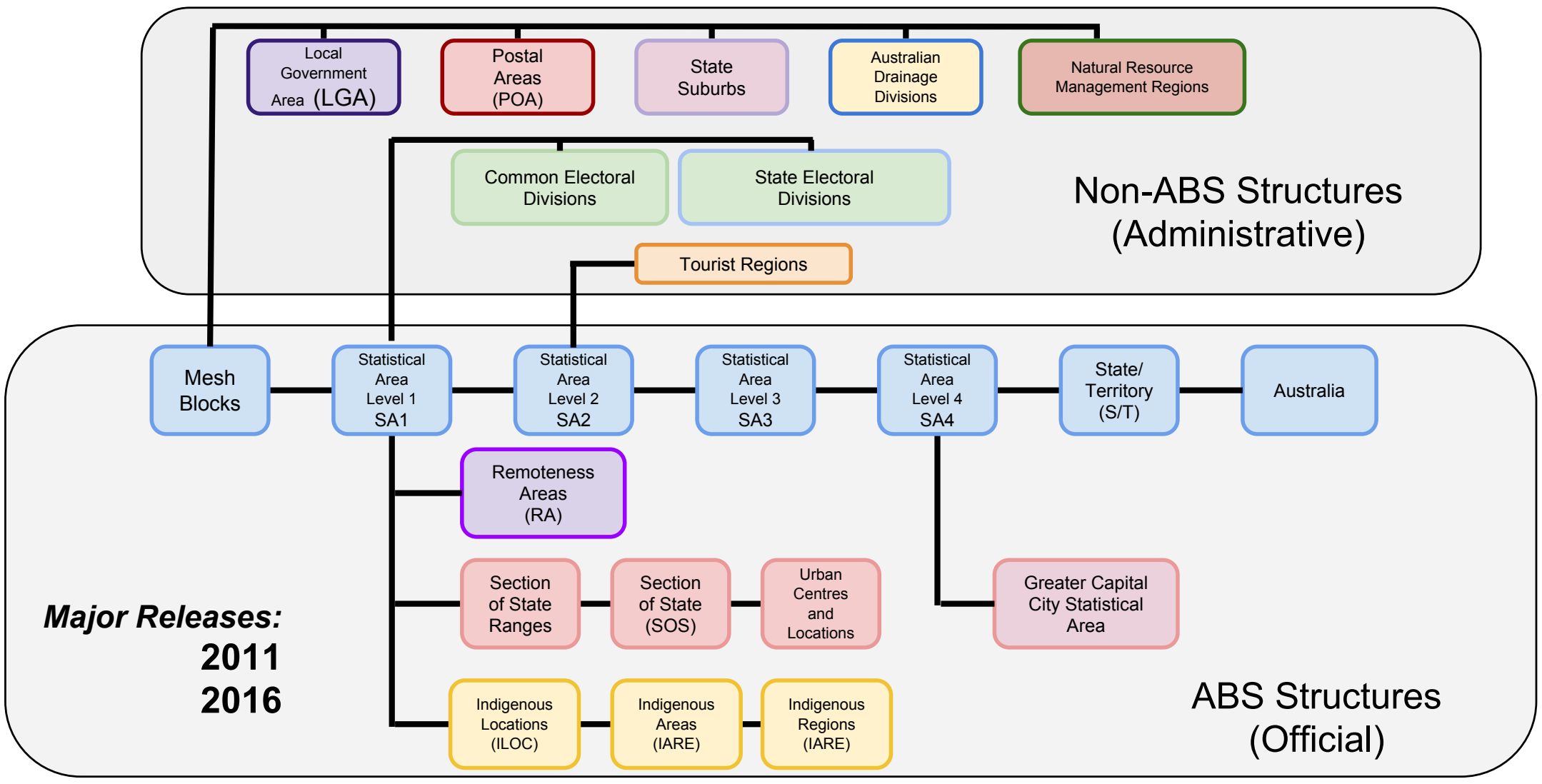

Figure 1: The ABS defines aggregations over a hierarchy of spatial polygon structures. The most fine-grained aggregation level is mesh block. Statistical Area level One regions (SA1s) are composed of mesh blocks, and all other ABS defined regions are composed of SA1s. Aggregations of ABS data to administrative structures such as electoral divisions are defined at the mesh block level.

There is therefore a high demand placed on the ABS for data *re-aggregation*. The ABS currently publish and maintain *correspondence* models for the most common re-aggregations, including converting from the 2011 ASGS boundaries to the 2016 ASGS boundaries, and between different resolution levels, such as SA2 to SA1. Currently, customers download *correspondence tables*, and apply them to the data manually. Many correspondences must be requested (such as Post Office Area to SA2), where the ABS will construct custom tables to provide to the data user. The ABS fields approximately 60 such requests per month.

Data61 is automating this process by building an accessible, open web service. The data user can access the service through a web HTTP interface that allows them to upload their original data, and download the re-aggregated data. This is itself is a significant contribution because it builds the model and applies it to the data automatically, automating the roles of both the ABS and the data user in this task. Given that modelling and data re-aggregation happen online, the capability of the service can be extended *per geometry* rather than *per correspondence*, and the model can be targeted to the subject of the user's data, where the current approach is restricted to static, pre-computed models.

Data61 has developed new re-aggregation algorithms for this service that offer multiple advantages over the current correspondences. Because re-aggregation is an ill-posed problem that admits a

**Key Terminology**

| | |
|---|---|
| $X_b$ | Supporting data at the *base geometry* aggregation, $n_b \times d$. |
| $Y_b$ | latent counts at the *base geometry* aggregation, $n_b \times 1$. |
| $Y_s$ | vector of counts at the *source geometry* aggregation, $n_s \times 1$. |
| $Y_d$ | vector of counts at the *destination geometry* aggregation, $n_d \times 1$. |
| $A_{sb}$ | Aggregation matrix relating *base geometry* counts to *source geometry* counts, $n_s \times n_b$ |
| $A_{db}$ | Aggregation matrix relating *base geometry* counts to *destination geometry* counts, $n_d \times n_b$ |
| $Z_b$ | latent parameters stochastically generating $Y_b$ (eg rates or probabilities of outcomes) at the *base geometry* aggregation, $n_b \times 1$. |

Figure 2: Terminology used throughout this document.

continuum of feasible solutions, the new algorithms quantify the uncertainty of their predictions over a multitude of possible solutions. This means that data users are given confidence intervals to support risk-aware decisions, where the existing approach provides a single *best guess* and cannot capture this inherent uncertainty.

Also, the new algorithms utilize machine learning to bring in additional related datasets, when available, to improve their accuracy. Instead of assuming a correlation with total population, the machine learning algorithm can use a rich set of related covariates, such as the census basic community profile, to build a predictive model of the user's data. For example, we could learn a predictive relationship between aged population counts in the census and a hospitalisation dataset, without knowing ahead of time that the user-data would be age correlated. The outcome is that, as well as providing confidence intervals, our new algorithms are expected to also achieve a higher predictive accuracy.

In the following sections we formalise aggregation (Section 2), polygon spatial geometry (Section 3) and the introduced related data (Section 4). We then present the (standard) population weighted algorithm in Section 5, and the (new) probabilistic algorithm in Section 6. A preliminary experimental validation is presented in Section 7, suggesting that the new algorithms do offer a predictive advantage, particularly in disaggregations where the predictions are higher resolution than the observations.

## 2 Aggregation

We formalise the concept of *aggregation* as a transformation from $n_b$ *base geometry* counts, $\mathbf{Y}_b$ , to $n_d$ *destination geometry* counts $\mathbf{Y}_d$ through a linear operator A:

$$Y_d = A_{db} Y_b \tag{1}$$

We additionally impose the following *aggregation* criteria, illustrated in Figure 3, to ensure the relation can be fully expressed by Equation 1:

**Definition 1** (complete allocation)**.** *Each source geometry element is fully covered by base geometry elements.*

**Definition 2** (unit allocation)**.** *Each base geometry element covers exactly one source geometry element.*

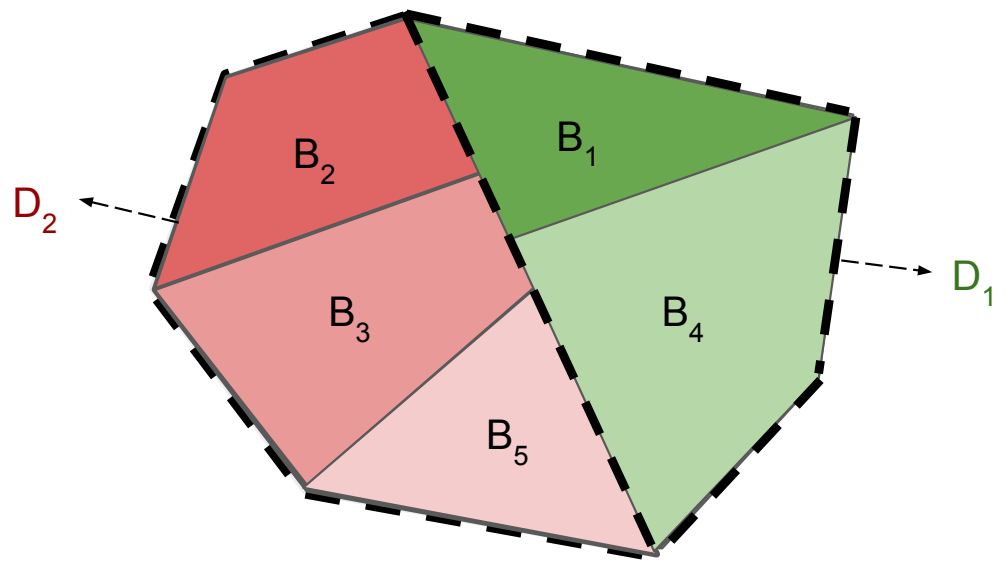

| $A_{db}$ | $B_1$ | $B_2$ | $B_3$ | $B_4$ | $B_5$ |
|---|---|---|---|---|---|
| $D_1$ | 1 | | | 1 | |
| $D_2$ | | 1 | 1 | | 1 |

Figure 3: Left: An aggregation correspondence is induced between counts over *base geometry* $B_1$, $B_2$, $B_3$, $B_4$, $B_5$ and counts over *destination geometry* $D_1$, $D_2$ due to common underlying point data. Right: The same relationship can be expressed as a sparse unit-entry aggregation matrix $A_{db}$ .

## 3 Base Geometry

In the re-aggregation problem, we actually have *two* aggregated vectors: a *source geometry* vector $Y_s$ that we observe, and a *destination geometry* vector $Y_d$ that we predict. The relation between $Y_s$ and $Y_d$ is generally more complicated than an *aggregation* because elements of the *source geometry* and elements of the *destination geometry* may partially overlap, violating the *unit allocation* criterion. The *actual* relation between $Y_s$ and $Y_d$ depends on the spatial location of the unit records - information that we don't have. Instead, without loss of generality, we can represent the underlying signal on a *base geometry* , carefully chosen so that both the $n_b$ observations $\mathbf{Y}_s$ and the $n_d$ predicted outputs $\mathbf{Y}_d$ are aggregations of $n_b$ latent counts $\mathbf{Y}_b$ . The convention of labeling the supporting geometry with subscripts b for *base*, s for *source* and d for *destination* is applied to all the terminology used in this document, as outlined in Figure 1. The full relation can then be expressed by *two* aggregations:

$$
\begin{aligned}
Y_s &= A_{sb} Y_b \\
Y_d &= A_{db} Y_b \qquad (2)
\end{aligned}
$$

The simplest admissible *base geometry* is given by the set of all intersections of a polygon $S \in$ *source geometry* with a polygon $D \in$ *destination geometry* . This geometric construction is depicted in Figure 4[Left]. However, given the hierarchical structure of the ABS statistical areas [3], an admissible base geometry can be found by identifying the first common ancestor in the composition hierarchy tree between *source geometry* and *destination geometry* as shown in Figure 4[Right]. As the diagram indicates, this will commonly be SA1 or meshblock.

## 4 Related Data

We have identified that the relation between $Y_s$ and $Y_d$ depends on the spatial distribution of the unit records (which we don't have). We therefore need to make a *model assumption*. Without knowing the subject of the aggregated counts ahead of time (the algorithm will be deployed as an automated service) we will require a general model suitable for use with a wide variety of possible problem subjects.

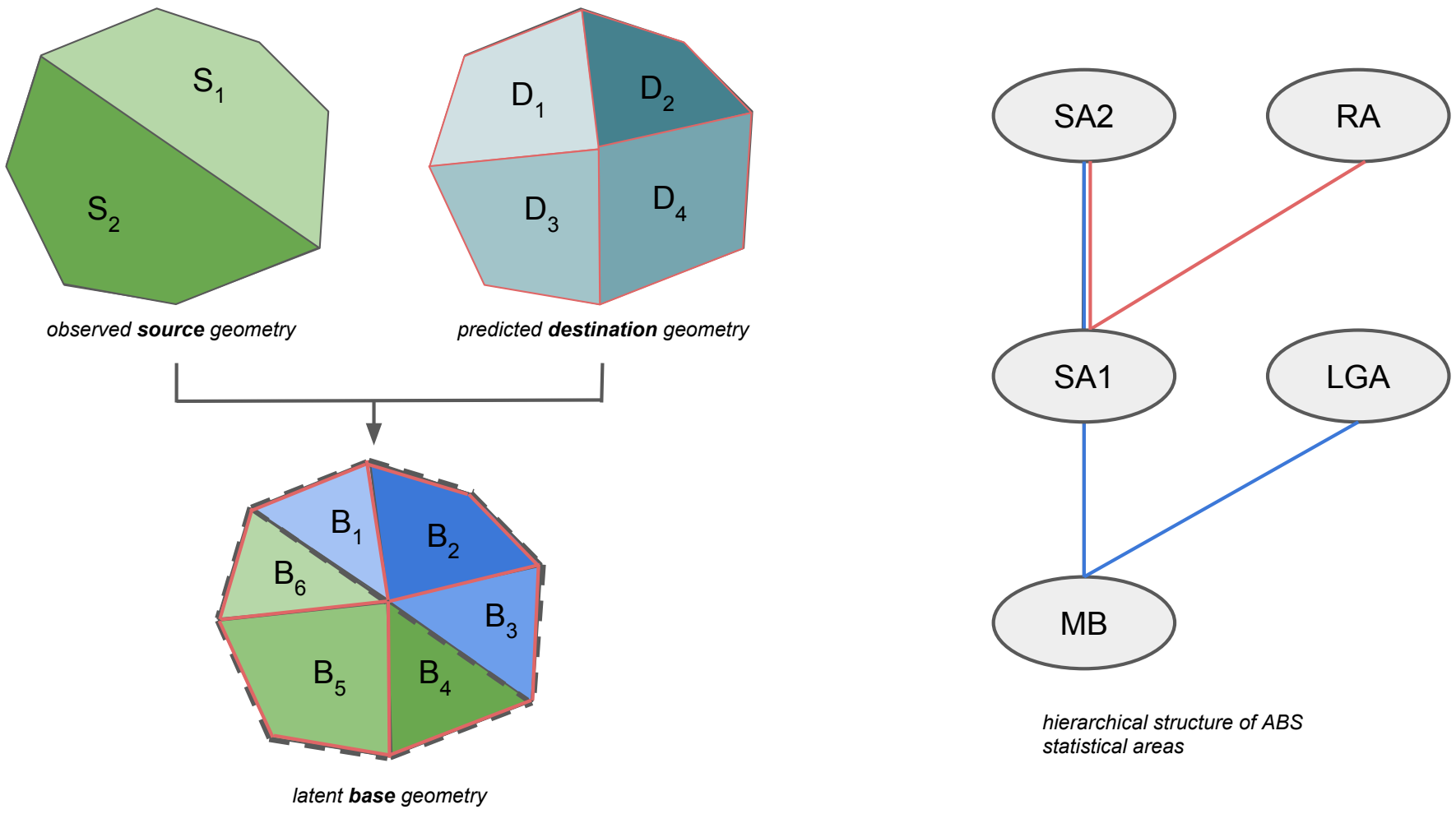

Figure 4: Left: Defining an *base geometry* for arbitrary *source geometry* and *destination geometry* in a spatial aggregation, we compute the set of all polygon intersections. Right: The hierarchical structure of the ABS statistical areas provides a natural set of base geometries determined by the first common ancestor in the structural tree. For example, a suitable *base geometry* for SA2 and RA is SA1 (shown by the red path), and SA2 to LGA is meshblock (the blue path).

The standard algorithm for re-aggregation is a *population weighted correspondence*, as outlined in Section 5. This algorithm assumes the correspondence is proportional to the number of common individuals shared by each *source geometry* and *destination geometry* . This assumption is encoded into a *correspondence table*, so for general use the ABS publishes correspondences calculated using related *total population* data.

On the other hand, our new *supervised machine learning* algorithms in Section 6 are able to automatically learn predictive statistical relationships between *any base geometry* data $X_b$ and the counts $Y_b$ without knowing the subject in advance. This means that instead of total population, the best general data for our machine learning approach is a rich general set of covariates with many dimensions that *may* be relevant.

For demographic re-aggregations, for example, a natural choice of *related data* will be census basic community profiles or agricultural census profiles aggregated at a SA1 or meshblock *base geometry* aggregation. We note that a *non-linear* predictive model might even benefit from having a *base geometry* at a higher resolution than the simplest admissible to resolve more detail.

We anticipate only two use-cases where a non-standard related dataset will be required. The first is when customers define their own aggregation polygons. In this case, the users would have to provide their geometry for analysis, for example by uploading shapefile records to the web API service. The second (more frequent) case arises when users update the *version* of their standard geometry, for example by re-aggregating from the 2011 ASGC boundaries to the 2016 ASGC boundaries. In either case, a finer sub-meshblock geometry will be geometrically defined as in Figure 4[Left]. We then require point data, such as G-NAF [16] or the ABS address register, to synthesise a basic population dataset using the point-polygon intersection strategy illustrated in Figure 5. The resulting population could then either serve *as* the new $X_b$ , or be used to re-aggregate the meshblock census covariates to the new *base geometry* with a hybrid population weighted approach.

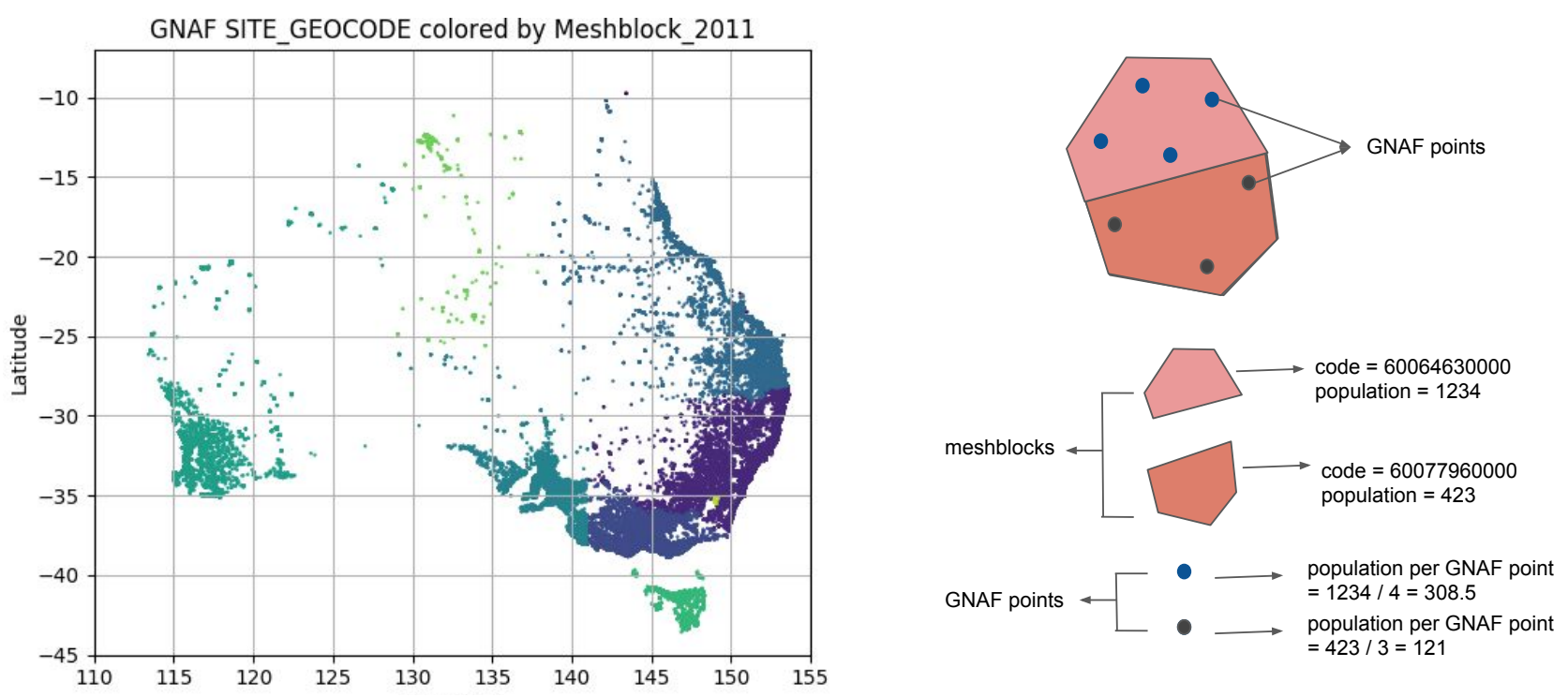

Figure 5: Related data will usually be taken from census publications. For non-standard geometry, population data can be synthesised by aggregating G-NAF or other address register point datasets over *base geometry polygons*. Left: co-registering Open G-NAF addresses with meshblock geometry (colouring addresses by meshblock index). Right: Efficient code has been written to intersect batches of point and polygon data scalable to national datasets.

## 5 Population Weighted Algorithm

*Population-weighted correspondences* are a linear re-aggregation model. They convert from $Y_s$ to a 'best guess' output $Y_d$ by allocating proportions of the *source geometry* observations to the *destination geometry* predictions based on the proprotion of common individuals. The model is encoded in the correspondence matrix $C_{ds}$ and is easily applied to new data through matrix multiplication:

$$Y_d = C_{ds} Y_s \tag{3}$$

Note the reversal of application direction compared to the *aggregation* matrices. Here we convert from observations to outputs, not from (latent) outputs to observations. We may derive a population weighted correspondence from $A_{sb}$ and $A_{sd}$ and $X_b$ as follows. We begin with an $X_b$ with one column, *total population*:

$$X_b^* : (n_b \times 1) \tag{4}$$

Although we can not do this with general use correspondence tables, sometimes the correspondence is targeted to a particular problem subject. For example, consider a *source geometry* and *destination geometry* that share a common retirement home population in their partial overlap. For counts of outcomes that are *statistically related to* the elderly demographic, such as *doctor home visits*, or *children per household*, the count correspondence will be much higher or lower than we would otherwise expect. In this case, $X_b^*$ may be derived from a sub-population dataset $X_b$ through a linear weighting:

$$X_b^* := X_b W \tag{5}$$

This is analogous to the weights *automatically learned* by the linear version of our predictive model in Section 6.3.1, except those weights can be learned online in response to new data.

We use the notation $\mathscr{D}_{X_b^*}$ to represent *expanding* the vector $X_b^*$ into a $n_b \times n_b$ diagonal matrix. The matrix recording the number of individuals in each *destination geometry* (by row) that lie in each *source geometry* (by column) is then given by:

$$P_{ds} = A_{db} \mathscr{D}_{X_b^*} A_{sb}^\top \tag{6}$$

The population in each *source geometry* region is given by the aggregation:

$$X_s = A_{sb} X_b^* \tag{7}$$

The observed rates of outcomes per individual is an element-wise ratio of $Y_s$ to $X_s$ that we can express as:

$$R_s = {\mathscr{D}_{X_s}}^{-1} Y_s = {\mathscr{D}_{A_{sb} X_b^*}}^{-1} Y_s \tag{8}$$

Now we bring in the *model assumption*: that the observed counts were uniformly distributed over the individuals in each *source geometry* . The *expected* number of counts in each *destination geometry* is therefore an accumulation of number of shared individuals multiplied by their outcomes-per-individual:

$$Y_d = P_{ds} R_s = A_{db} \mathscr{D}_{X_b^*} A_{sb}^\top {\mathscr{D}_{A_{sb} X_b^*}}^{-1} Y_s \tag{9}$$

We note that computing single value here leads to a 'best guess' prediction. The advantage is that this model is easily stored as a matrix, and can be readily applied to *any* $Y_s$ using Equation 3, with the equivalent correspondence matrix:

$$C_{ds} = A_{db} \mathscr{D}_{X_b^*} A_{sb}^\top {\mathscr{D}_{A_{sb} X_b^*}}^{-1} \tag{10}$$

The disadvantage of using a pre-prepared $C_{ds}$ is that it only generates the 'best guess' prediction and we are disregarding the uncertainty associated with the model assumption and the process noise. Ideally we would address this using a distribution of $\mathbf{C_{ds}}$. This motivation, together with the desire to automatically select a model for any problem subject, has led to the development of new probabilstic machine learning algorithms below.

# 6 Probabilistic Machine Learning Algorithm

Contrasting the population correspondence above, our new algorithms interpret $A_{sb}$ and $A_{db}$ as *forward* sensor models because each is transforming the common underlying signal $Y_b$ to a new representation. This frames the re-aggregation problem within the class of *sensor inversion problems* faced in many physical sensing systems, analogous to determining the rock density field underlying a geophysical gravity survey, or the 3D brain structure underlying raw tomography data [2]. Our strategy is therefore to *invert* the observations into a *distribution* over of predictions consistent with the observations. The model is constructed with the following *generative* story:

**Definition 3** (population)**.** *the set of people/households/addresses in the experiment.*

**Definition 4** (experiment)**.** *the interaction with the population that produced the unit records, such as recording hospitalisations, or asking a survey question.*

$P(\mathbf{Y}_b|X_b)$ is the generative distribution of *this* experiment's outcomes when applied to *any* population with features $X_b$ . We assume that if we repeated the experiment on other distinct populations with the same $X_b$ then each instance $Y_b$ would be a draw from this distribution.

$P(\mathbf{Y}_s = Y_s|Y_b, A_{sb})$ is the observational likelihood that *a* draw from $\mathbf{Y}_b$ would aggregate to the *observed* $Y_s$ .

The probabilistic graphical model corresponding to this view has inputs $X_b$ , $\theta$ , $A_{sb}$ and $A_{db}$ . The joint distribution factorisation as shown in Figure 6 is given by:

$$P(\mathbf{Y}_s, \mathbf{Y}_d, \mathbf{Y}_b, \mathbf{Z}_b|X_b, \theta, A_{sb}, A_{db}) = P(\mathbf{Y}_s|A_{sb}Y_b)\ P(\mathbf{Y}_d|A_{db}Y_b)\ P(\mathbf{Y}_b|Z_b)\ P(\mathbf{Z}_b|X_b, \theta) \tag{11}$$

The bold nodes $\mathbf{Z}_b$ , $\mathbf{Y}_b$ , $\mathbf{Y}_s$ and $\mathbf{Y}_d$ are the above *conditional distributions*, interpreted in the subsections below.

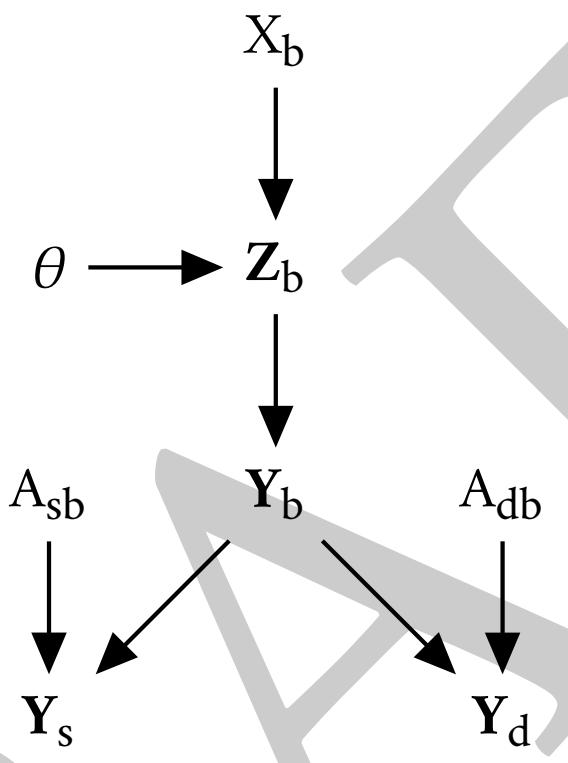

Figure 6: Probabilistic generative model $P(\mathbf{Y}_s, \mathbf{Y}_d, \mathbf{Y}_b, \mathbf{Z}_b|X_b, \theta, A_{sb}, A_{db})$.

## 6.1 Process Likelihood $P(\mathbf{Y}_b|Z_b)$

We interpret $P(\mathbf{Y}_b|Z_b)$ as the probability of stochastically generating discrete count outcomes $\mathbf{Y}_b$ from the continuous process parameters $Z_b$ . The selection of this distribution is informed by the type of *experiment*:

**event** data aggregates independent random events. $Z_b$ is a vector of positive real *rates* within each geometry, such that independently $\mathbf{Y}_b \sim \text{Poisson}(Z_b)$. For example, if $Y_s$ is *counts* of hospitalisation at the *base geometry* , then $Z_b$ might be latent *rates* of hospitalisation with a Poisson process likelihood.

**vote** data aggregates exactly one Bernoulli outcome per [person/unit area] within each geometry. $Z_b$ is a vector of latent outcome probabilities, $0 \le z_i \le 1$, such that $\mathbf{Y}_b \sim \text{Binomial}(Z_b)$. For example, if $Y_s$ is *counts* of 'yes' answers to a yes/no question at the *base geometry* , then $Z_b$ might be latent *probabilities* of answering 'yes' with a Poisson process likelihood.

**ordinal** data contains arbitrary attributes averaged over region populations. Often a Gaussian likelihood will be a good choice as it is highly tractable to work with. For example, if $Y_s$ is mean income of individuals in the *base geometry* , then $Z_b$ might remain income with a Gaussian likelihood to capture variability.

### 6.2 Aggregations $P(\mathbf{Y}_s|Y_b)$ and $P(\mathbf{Y}_d|Y_b)$

In every realisation of the model, $Y_s$ and $Y_d$ are aggregations of a common signal $Y_b$ as specified in Equations 1 and 2.

### 6.3 Generative Model $P(\mathbf{Z}_b|X_b, \theta)$

We have factorised $P(\mathbf{Y}_b|X_b)$ to include process parameters $\theta$ , and a continuous latent function $\mathbf{Z}_b$ :

$$P(\mathbf{Y}_b|X_b) = P(\mathbf{Y}_b|Z_b)\, P(\mathbf{Z}_b|X_b, \theta) \tag{12}$$

We have already assumed a form for $P(\mathbf{Y}_b|Z_b)$. We now apply a machine learning approach to $P(\mathbf{Z}_b|X_b)$ by defining an expressive parameterised function:

$$\mathscr{F}(X_b, \theta) \rightarrow P(Z_b) \tag{13}$$

We then use *supervised machine learning* to infer its parameters $\theta$ , together with any model *latent* parameters, based on the *examples* $(X_b, Y_s)$. Note that the predictive $Y_d$ and $A_{db}$ do not feature at all in the learning (model selection) process. Below we outline two model candidates that are being examined:

#### 6.3.1 Linear Regressor

A simple model structure is a *linear* relation between $X_b$ and $Z_b$ :

$$Z_b = X_b W \tag{14}$$

Here, we introduce $W$ , our $d \times 1$ *weight vector* where d is the number of *related* dimensions. These weights have an equivalent role to the *weighting factor* used to select sub-populations in a population-weighted approach, but they are learned automatically. A simple approach is to select the *maximum-likelihood weights* W through numerical optimisation.

However, a single weight vector does not quantify the model uncertainty relating to $X_b$ . A Bayesian approach can better capture *model uncertainty* by maintaining a distribution $\mathbf{W}$

$$\mathbf{Z}_b = X_b \mathbf{W} \tag{15}$$

#### 6.3.2 Deep Neural Network Regressor

Deep networks are state of the art machine learning models based on composing layers of linear models with non-linear *activation* functions [9]. A Bayesian view of this basic structure is illustrated in Figure 7. In this structure we have k sequential linear projections with probabilistic weights $\mathbf{W}_0$, $\mathbf{W}_1 \ldots \mathbf{W}_k$. The outputs of all but the last layer are *latent*, and we therefore *design* both the number of layers and the number of nodes in each layer by choosing the dimensionality of each $\mathbf{W}$, keeping in mind we must learn *weights* on the connections. Learning (and specifying) these models is a challenging and very active problem in the machine learning literature [8]. Data61 has published an open source library for constructing and learning deep Bayesian models [7].

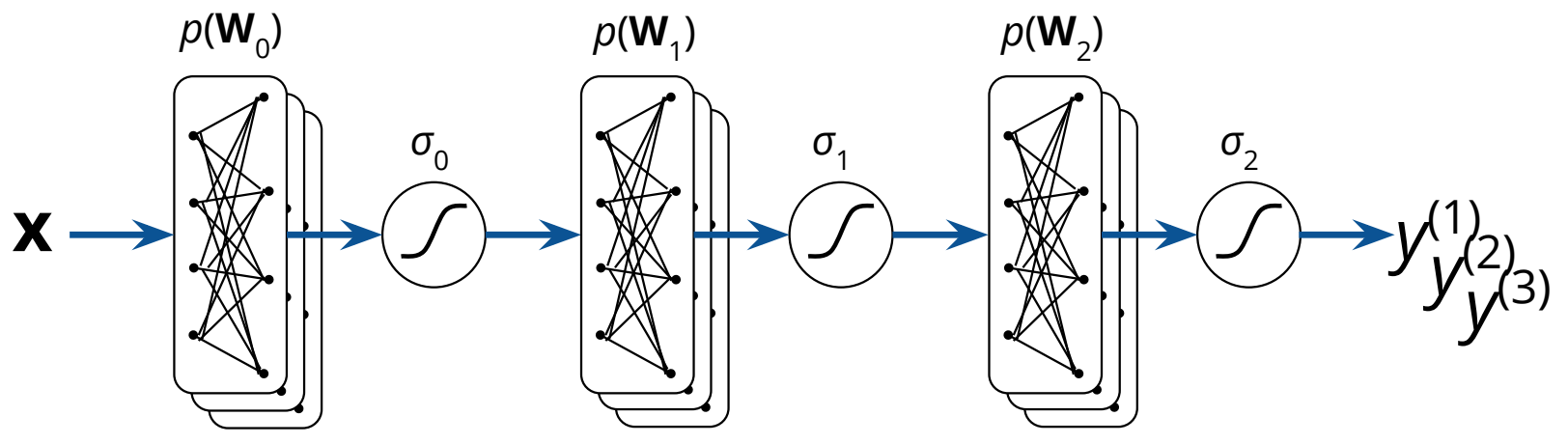

Figure 7: Structure of a Bayesian deep network with two latent (hidden) layers and one output layer. During learning, a stochastic variational approach is used to determine the hyper-parameters of the weight distributions. Joint predictions may be sampled from the model to obtain expectations and confidence intervals. Given the stochastic formulation, we query this model by sampling joint outcomes $y^{(i)} \sim \mathbf{Y}$

The resulting framework is a highly expressive, flexible machine learning model that can learn from data and quantify its predictive uncertainty. This flexibility comes at the cost that the models require much more computing resources than simpler linear models and may not be suitable for deployment on a government-hosted service.

## 6.4 Predictive Inference

Inference on this model involves four steps:

**learning** the model parameters $\theta$ and latent parameters W using supervised machine learning

**inferring** the joint generative distribution $P(\mathbf{Y}_s, \mathbf{Y}_d, \mathbf{Y}_b, \mathbf{Z}_b | X_b, \theta, A_{sb}, A_{db})$ for *any* population with features $X_b$ by querying the above model on the inputs $X_b$ , $A_{sb}$ , $A_{db}$ and $\theta$ .

**conditioning** the above distribution on $\mathbf{Y}_s = Y_s$, to extract the sub-distribution that is *also* consistent with the observations

**summarising** the marginals of $\mathbf{Y}_d$ to extract predictive expectations and confidence intervals.

### 6.4.1 Learning

One strategy to select the latent model weights is to learn the maximum-a-posteriori model by *optimising* W to fit the aggregated observations:

**Definition 5** (Maximum a-Posteriori (MAP) model selection)**.**

$$W = \underset{W^*}{\text{argmax}} \; P(\mathbf{Y}_s = Y_s | Y_b, A_{sb}) \, P(Y_b | Z_b) \, P(Z_b | X_b, W) \, P(W^* | \theta) \tag{16}$$

Here the role of the prior $P(W^*|\theta)$ is to penalise overfitting, as it is crucial we capture a general relation between $X_b$ and $Y_s$ , and don't memorise the data. We may even learn the prior by optimising $\theta$ , although this has to be done carefully with respect to performance on hold-out data unseen by the weight selection mechanism.

We then introduce a prior on W to prevent over-fitting due to highly specific weights. Here $\theta$ contains one parameter, $\lambda$, that specifies a regularisation on the choice of W , such as a spherical Gaussian prior:

$$P(W|\lambda) \sim \mathcal{N}(\mathbf{W}|0, \lambda^{-1}I) \tag{17}$$

To quantify the model uncertainty relating to $X_b$ , we take a Bayesian approach and maintain a distribution $\mathbf{W}$ instead of selecting the *maximum-a-posteriori* W . The hyper-parameters $\theta$ can then be learned by marginalising over $\mathbf{W}$ , leading to a *type 2 marginal likelihood* model selection:

**Definition 6** (Type 2 marginal likelihood model selection)**.**

$$\Theta = \underset{\Theta^*}{\text{argmax}} \int P(\mathbf{Y}_s | \mathbf{Y}_b, A_{sb}) \, P(\mathbf{Y}_b | \mathbf{Z}_b) \, P(\mathbf{Z}_b | X_b, \mathbf{W}) \, P(\mathbf{W} | \Theta^*) \, P(\Theta^*) \, \delta \mathbf{W} \tag{18}$$

This criterion is naturally good at ensuring the resulting model will generalise, as it penalises both model error and model complexity. The disadvantage is that, unless all the factors are conjugate, the integral will be intractable. Sampling and stochastic optimisation is generally required, and this is handled by Data61's deep learning library for both the the Bayesian linear model and the Bayesian deep-net model using a stochastic variational approximation to the *type 2 marginal likelihood* from the literature [10] [6].

Machine learning approaches rely on a statistical relationship (not necessarily correlation) between $\mathbf{Y}_b$ and $X_b$ . The REST API service provides a heuristic to test this assumption based on hold-out prediction of $\mathbf{Y}_s$ . We note that this analysis is not a validation of the re-aggregation, as we wont have ground truth at deployment. The diagnostic instead compares the cross validation predictive density of a Bayesian linear model against the predictive density of the Gaussian marginal of $\mathbf{Y}_s$ . Thus we can detect when a model using ($X_b$ , $\mathbf{Y}_s$ ) does no better than one only using $\mathbf{Y}_s$ and the dependence is weak.

### 6.4.2 Inference

We can query the model for particular inputs $X_b$ , $\theta$ , $A_{sb}$ , $A_{db}$ by propagating the conditional probability distributions through ($\mathbf{Z}_b, \mathbf{Y}_b, \mathbf{Y}_s, \mathbf{Y}_d$). An unusual aspect of this formulation, compared to most machine learning models, is that we apply aggregation transformations *after* the model likelihood. This is neccessary for $Y_s$ and $Y_d$ to share a common signal, but integrating over $\mathbf{Y}_b$ to compute the observational likelihood $P(\mathbf{Y}_s = Y_s|Y_b)$ can lead to computational problems.

We identify some distributions for which we can use *closed form aggregation* to make evaluating $P(\mathbf{Y}_s = Y_s|Y_b)$ tractable.

**Gaussian Likelihoods** : If $Y_b$ is multivariate Gaussian with mean $M$ and covariance $\Sigma$:

$$\mathbf{Y}_b \sim \mathcal{N}(M, \Sigma) \tag{19}$$

then the distribution over $\mathbf{Y}_s$ is a closed form expression:

$$P(A_{sb}Y_b) = \mathbf{Y}_s \sim \mathcal{N}(A_{sb}M, A_{sb}\Sigma A_{sb}^{\top}) \tag{20}$$

**Poisson Likelihoods** : If $Y_b$ is *iid* Poisson distributed with rates $R_b$:

$$\mathbf{Y}_b \sim \text{Poisson}(R_b) \tag{21}$$

then the distribution over $\mathbf{Y}_s$ is a closed form expression:

$$P(A_{sb}Y_b) = \mathbf{Y}_s \sim \text{Poisson}(A_{sb}R_b) \tag{22}$$

**Binomial Likelihoods** : If $Y_b$ is *iid* Binomial distributed with probability P and population N:

$$\mathbf{Y}_b \sim \text{Binomial}(N_b, P_b) \tag{23}$$

Then $P(A_{sb}Y_b)$ is a computationally intractable *sum of binomials* distribution. Possible strategies for using this likelihood include *sampling* or *approximation*. For example, if we optimise $P_b^*$ to match the true expectation, the following will approximate the likelilihood (but under-estimate the variation).

$$P(A_{sb}Y_b) = \mathbf{Y}_s \approx \text{Binomial}(A_{sb}N_b, P_b^*) \tag{24}$$

### 6.4.3 Conditioning

Now that we have a generative model, we extract its latent distribution $P(\mathbf{Y}_b|X_b)$. Through *aggregation* of common instantiations $Y_b$ , this implies a joint distribution $(\mathbf{Y}_s, \mathbf{Y}_d)$. We want to select the component of this distribution that is consistent with our observations.

This means we need to *invert* the observations: we have $n_s$ observations constraining the $n_b$ latent counts, leaving $n_f = n_b - n - s$ degrees of freedom. The appropriate linear algebra tool to parameterise the solutions is the *null space* of $A_{sb}$ .

**Definition 7.** *the null space of* $A_{sb}$ *is the* $n_b \times n_f$ *linear mapping* $N_{bsf}$ *between vector spaces* $Y_b$ *and* $Y_s$ *such that for any* $n_f$ *vector* $V_f$*:*

$$\forall V_f \quad A_{sb}N_{bsf}V_f = \mathbf{0} \tag{25}$$

We can think of $V_f$ as vector of 'free' coefficients parametrising the position on the solution plane, and $N_{bsf}$ as the corresponding co-ordinate system unit vectors on the hyperplane. There are standard algorithms for finding $N_{bsf}$ based on matrix row reduction [11]. Then we identify *any* candidate solution $\bar{Y}_b$ such that:

$$A_{sb}\bar{Y}_b = Y_s \tag{26}$$

For example, the pseudo-inverse yields one solution:

$$\bar{Y}_b = A_{sb}^\top (A_{sb} A_{sb}^\top)^{-1} Y_s \tag{27}$$

Then the space of *all* candidates can be parameterised on the hyper-plane as a function of $V_f$ :

$$Y_b := \bar{Y}_b + N_{bsf} V_f \tag{28}$$

We are now faced with the computational task of slicing an $n_b$ dimensional probability distribution along this $n_f$ dimensional hyper-plane. This is a highly non-trivial problem for general distributions. Constraints such as *positivity* will also restrict the valid domain. These constraints can either be projected into the null space, or we may define a mapping between an unbounded continuous latent space and positive $\mathbf{Y}_b$ . Because the algorithm will be deployed on a government server, we are currently exploring possible algorithms that offer different trade offs in terms of scalability and precision. Three potential strategies are identified below and illustrated with the following toy problem.

**Conditioning: Toy problem**

The following strategies are illustrated in Figure 8 using a compact toy problem. The details of this problem are as follows. We have two base regions:

$$\mathbf{Y}_b = \begin{bmatrix} y_1 \\ y_2 \end{bmatrix} \tag{29}$$

We have the trivial aggregation relation:

$$\mathbf{Y}_s = y_1 + y_2 \qquad A_{sb} = \begin{bmatrix} 1 & 1 \end{bmatrix} \tag{30}$$

And one observation of the single element $\mathbf{Y}_s$ :

$$Y_s = 100 \tag{31}$$

We are inferring a two dimensional count vector, from one observation, so the latent space is one dimensional (a line). We assume positive counts, so the valid line interval is bounded to the first quadrant. Note that in general this space will be a $n_f$ dimensional hyperplane with $n_b$ constraints, which is why we need to explore tractable computation. In this toy problem, the valid latent space is the locus:

$$y_1 + y_2 = 100 \qquad y_1 \geq 0, y_2 \geq 0 \tag{32}$$

We have [arbitrary] side data $X_b$ at the $X_b$ aggregation with many feature columns:

$$X_b = \begin{bmatrix} x_{11} & x_{12} & \dots \\ x_{21} & x_{22} & \dots \end{bmatrix} \tag{33}$$

And query a trained machine learning model to obtain a predictive *distribution* over how regions with these features might expect outcomes to occur. The distribution depends on the likelihood selection and the type of model, but for simple illustrative purposes a Gaussian is used in this example:

$$P(\mathbf{Y}_b | X_b) = f(X_b) \sim \mathcal{N}\left(\begin{bmatrix} 50 \\ 35 \end{bmatrix}, \begin{bmatrix} 200 & 0 \\ 0 & 100 \end{bmatrix}\right) \tag{34}$$

The applications of the three following computational strategies on this toy problem are depicted in Figure 8.

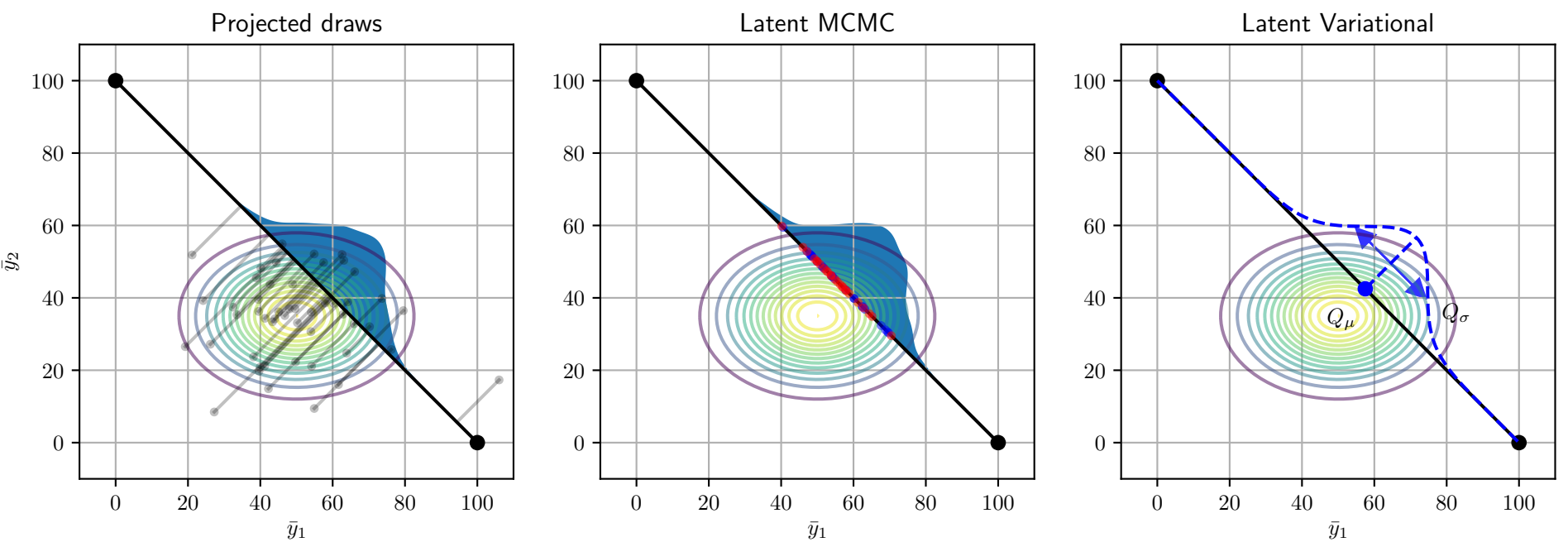

Figure 8: Illustrating the three suggested conditioning strategies on the toy problem. The plot axes represent the space of $\mathbf{Y}_b$ . The black line represents solutions in $\mathbf{Y}_b$ consistent with $\mathbf{Y}_s$ . The countours show $P(\mathbf{Y}_b|X_b)$. The blue curves show the inferred probability density along the line. [Left] samples from the model are projected to the line. [Center] MCMC samples on the line directly (with accepts and rejects shown as red and blue dots). [Right] Variational MCMC places a surrogate distribution $\mathbf{Q}$ on the line, and minimises the divergence between $\mathbf{Q}$ and $P(\mathbf{Y}_b|X_b)$ in the space of $\mathbf{Y}_b$ by optimising parameters $Q_\mu$ and $Q_\sigma$ with a tractable lower bound.

### Conditioning strategy: projected samples

This strategy draws samples from $P(\mathbf{Y}_b|X_b)$. We then make a computationally inexpensive, low precision approximation. Ideally we would reject samples that to not aggregate, but the probability drawing a valid high dimensional sample naively is essentially zero. Instead, we *project* each sample to its nearest point on the solution domain using the null space re-parameterisation of Equation 28:

$$Y_b^* = \bar{Y}_b + (Y_b - \bar{Y}_b)N_{bsf}N_{bsf}^\top \tag{35}$$

We may then use the statistics (quantiles, expectation) of the projected samples as an approximation of the exact statistics.

### Conditioning strategy: null space MCMC

We may attempt to sample from the solution space directly using Markov Chain Monte Carlo (MCMC) [13]. This approach random-walks the probability distribution, using each previous sample as a basis for proposing the next. This method is potentially very costly because it involves proposing (and ultimately rejecting) a large number of samples.

We can efficiently explore the posterior using the re-parameterisation in Equation 28, accepting or rejected samples according to $P(Y_b|X_b)$ where $Y_b = \bar{Y}_b + N_{bsf}V_f$. Even with this parameterisation, we still have restricted domain for count data where each $\mathbf{Y}_{b_i} \geq 0$ that may lead to sampling difficulty. Recent developments in this field that may increase the sampling efficiency include *Hamiltonian* MCMC where a particle simulation using the gradient of the probability density is used to compute proposals [12], and *constrained* Hamiltonian simulation where the proposal path may reflect off constraints in manner that does not require the sample to be rejected [4].

**Conditioning strategy: latent variational model**

Instead of exploring the posterior through sampling $\mathbf{V}_f$ , we could define a simpler parameterised distribution $\mathbf{Q}$ in the null space (such as a Gaussian) and infer its parameters. We then seek to minimise the difference (in $Y_b$ space) between the surrogate distribution and the generative model distribution by *optimising the parameters of the surrogate distribution*. Variational inference provides an evidence lower bound criterion to make this optimisation much more tractable than the MCMC approach [14], but potentially more accurate than the projection approach. Once we have learned the parameters of $\mathbf{Q}$ , we can evaluate statistics of $\mathbf{Y}_b$ ($\mathbf{Q}$ ) easily with sampling or integration due to the simpler form of the surrogate.

### 6.4.4 Summarising

The conditioning strategy gives us a distribution $\mathbf{Y}_b$ . We simply need to aggregate this to our predicted modality $\mathbf{Y}_d$ . This could either be in closed form, or through sampling. Currently we present the result to the user as statistics of the marginals, specifically by providing the expectation, lower and upper quantiles.

## 7 Preliminary Validation Experiments

To validate a re-aggregation algorithm, we require datasets created from the same unit records at two or more aggregation levels. Related studies of correspondence accuracy (without confidence intervals ) have been conducted within the ABS [15]. Many suitable validation datasets are available relating to the Australian population and land use. To date, several have been used to test the performance of the re-aggregation algorithm, including the ABS publication *Statistics for Deaths and Mortality in Australia* for 2011 [1], where aggregations to SA4, SA3, and SA2 are available.

Figure 9 compares the performance of the *new algorithm* with a baseline *population weighted* correspondence at different levels of reaggregation. Note that neither method knew in advance that the targets would be deaths. Therefore the population weighted method uses *total population*, and the Bayesian method is learning *online* from the SA1 basic community profile. Both algorithms are Data61's prototype implementations using the same population data. Importantly, the new algorithm has added *per-region* confidence intervals (shown as red bars), where calibrated uncertainty is reflected by the intervals reaching the diagonal and thus containing the true target value.

We apply standard metrics to quantify predictive accuracy. Two such metrics appropriate for comparing point 'best guess' estimates are R2 and RMSE:

**R**2 **(R-squared)** is a normalised statistical measure of how close the predicted values are to the true values, where scores closer to 1.0 indicate better performance.

**RMSE (Root-Mean-Squared Error)** is a measure of the magnitude of error between the predicted and true values, where closer to zero indicates better performance.

Probabilistic results also allow for predictive *density* performance measures that penalise under and over-confidence as well as accuracy. These metrics will only be appropriate for comparing two or more *probabilistic* models, and can not be used in this comparison. However, they will be used in

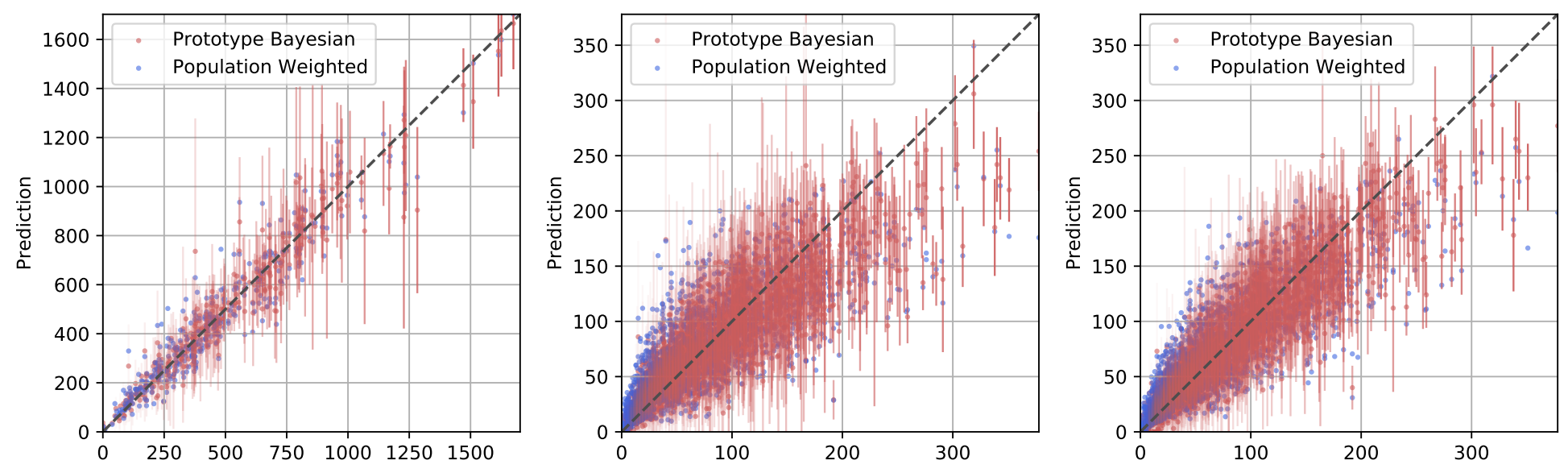

Figure 9: Comparison of the new development prototype model (red with confidence intervals) against a population weighted correspondence (blue point estimates). Each axis provides a scatter plot where the x-axis represents true death count while the y-axis shows predicted death count. Three re-aggregations were applied to the Death count dataset for increasingly difficult re-aggregations: [Left] SA4(2011) to SA3(2011), [Centre] SA3(2011) to SA2(2011), [Right] SA4(2011) to SA2(2011),

future experiments to inform model design and development. A popular metric is NLP (Negative Log Probability):

**NLP (Negative Log Probability)** directly measures the predictive probability density placed on the true values. Smaller is better, but this metric has no absolute interpretation and can only be interpreted in relation to another NLP score.

With the caveat that these are preliminary findings and the model is under active development, the performance metrics corresponding to Figure 9 are given in Table 1 and Table 2. As well as adding confidence intervals to the output, the initial validation indicates that the new algorithm also offers higher accuracy than a total-population weighted prediction. We expect a predictive advantage because machine learning algorithms are able to automatically learn a new demographically dependent function that could, for example, draw upon age information in this death context. This advantage is found to grow as the detail of the prediction geometry increases relative to the observation geometry. This is because the observations admit an increasingly wide range of solutions and so considering the *space* of feasible solutions and bringing in rich related data becomes critical in making detailed predictions. Ongoing quantitative validation will be conducted as the model is further developed.

| | SA4-SA3 | SA3-SA2 | SA4-SA2 |
|---|---|---|---|
| New Algorithm (Prototype) | 0.94 | 0.86 | 0.75 |
| Population Weighted | 0.92 | 0.73 | 0.66 |

Table 1: R2 metrics comparing the new probabilistic algorithn with a population weighted algorithm on the deaths dataset at three aggregations (closer to 1.0 is better).

| | SA4-SA3 | SA3-SA2 | SA4-SA2 |
|---|---|---|---|
| New Algorithm (Prototype) | 5909 | 482 | 846 |
| Population Weighted | 7774 | 913 | 1143 |

Table 2: SMSE metrics comparing the new probabilistic algorithn with a population weighted algorithm on the deaths dataset at three aggregations (closer to 0 is better, and this metric is sensitive to the size of the testing dataset).

## References

[1] ABS. *Deaths, Australia (Data Cube)*. 2016. URL: http://www.abs.gov.au/AUSSTATS/abs@.nsf/DetailsPage/3302.02016?OpenDocument.

[2] Richard Aster, Brian Borchers, and Clifford Thurber. *Parameter Estimation and Inverse Problems*. 2nd ed. Elsevier, 2012.

[3] *Australian Statistical Geography Standard (ASGS)*. 2011, 2016.

[4] M Betancourt. "Nested Sampling with Constrained Hamiltonian Monte Carlo". In: *AIP Conference Proceedings* 165 (2011).

[5] Christopher M. Bishop. *Pattern recognition and machine learning*. Vol. 1. Springer New York, 2006.

[6] Kurt Cutajar et al. "Random Feature Expansions for Deep Gaussian Processes". In: *International Conference on Machine Learning*. 2017.

[7] CSIRO Data61. *Aboleth Bayesian Deep Learning Library*. 2017. URL: https://github.com/data61/aboleth.

[8] Ian Goodfellow, Yoshua Bengio, and Aaron Courville. *Deep Learning*. http://www.deeplearningbook.org. MIT Press, 2016.

[9] Geoffrey E. Hinton and Simon Osindero. "A fast learning algorithm for deep belief nets". In: *Neural Computation* 18 (2006), p. 2006.

[10] Diederik P Kingma and Max Welling. "Auto-Encoding Variational Bayes". In: *International Conference on Learning Representations*. 2014.

[11] David Lay. *Linear Algebra and its Applications*. 3rd ed. Addison Wesley, 2005.

[12] Radford M. Neal. "MCMC Using Hamiltonian Dynamics". In: *Handbook of Markov Chain Monte Carlo* 54 (2010), pp. 113–162.

[13] Radforn M. Neal. *Probabilistic Inference Using Markov Chain Monte Carlo Methods*. Tech. rep. CRG-TR-93-1. Dept of Computer Science, University of Toronto, 1993.

[14] Manfred Opper and Cedric Archambeau. "The variational Gaussian approximation revisited". In: *Neural Computation* 21 (3 2009), pp. 786–792.

[15] Brian Pink. *Converting Data to the Australian Statistical Geography Standard*. Tech. rep. Australian Bureau of Statistics, 2012.

[16] *PSMA Geocoded National Address File (G-NAF)*. 2016.